\documentstyle[epsfig,twocolumn]{elsart}
\begin{document}
\begin{frontmatter}

\title{Depairing currents in superconductor / ferromagnet Nb/CuNi trilayers close to T$_c$.}
\author{A.Rusanov},
\author{M.Hesselberth},
\author{S.Habraken} and
\author{J.Aarts}

\address{Kamerlingh Onnes Laboratory, Leiden University, P.O. Box 9506,
2300 RA Leiden, The Netherlands}

\date{17-oct-03}

\begin{abstract}
In superconductor/ferromagnet (S/F) heterostructures, the exchange field $h_{ex}$ of the
F-layer suppresses the superconducting order parameter in the S-layer via the proximity
effect. One issue in current research is the effect of a domain state, or, more
generally, different directions of $h_{ex}$, on the superconductivity. We used a
pulsed-current technique in order to measure the superconducting transport properties of
Cu$_{1-x}$Ni$_x$/Nb/Cu$_{1-x}$Ni$_x$ ($x$ = 0.54) F/S/F trilayers structured in strips of
about 2 $\mu$m wide and 20 $\mu$m long as function of a small in-plane magnetic field. We
find that the depairing current is tied to the magnetization behavior. In particular, we
show that the suppression of superconductivity in the S-layer is smallest when the
external magnetic field equals the coercive field H$_{c}$ of the F-layers.
\\
$\newline$ Keywords: Proximity effect, FSF-junctions, FS heterostructures, spintronics.
\end{abstract}

\end{frontmatter}

\small

\section{Introduction}
The issue of the proximity effect between a superconductor and a ferromagnet is the focus
of much current research. In principle, the superconductivity in the S layer is
suppressed due to the pair breaking of the Cooper pair by the exchange field $h_{ex}$
experienced in the ferromagnet. In the F-layer, a superconducting order parameter can
still exist, but it becomes spatially modulated because the electrons which create a
Cooper pair belong to different spin subbands. This leads to several effects observed
experimentally, including oscillatory behavior of the superconducting transition
temperature T${_c}$ as function of ferromagnetic layer thickness in F/S bi- or
multilayers \cite{rado91,lazar00,garif02}, and $\pi$-junctions in S/F/S Josephson
junctions \cite{ryaz01,guich03}. A different mechanism to influence the superconducting
order parameter in the S-layer by the adjacent F-layers is by varying the relative
directions of the magnetization (and therefore $h_{ex}$) in the two F-layers. It was
predicted that, when the thickness of superconductor d$_S$ is of the order of the
superconducting coherence length $\xi_S$, the superconducting transition temperature is
higher when the two magnetization directions are antiparallel than when they are parallel
to each other \cite{buzd99}. The difference can even be enhanced by tuning the thickness
of the F-layers $d_F$ such that $d_F \approx$ $\xi_F/2$ (with $\xi_F$ the coherence
length in the ferromagnet) \cite{tagir99}. A first observation of these so-called
spin-switch effects was recently reported in ref.~\cite{gu02}. Basically, spin switching
comes about when the Cooper pair samples different directions of $h_{ex}$ simultaneously,
which raises the question whether similar effects can be observed if the F-layers contain
a domain structure rather than a homogeneous magnetization~\cite{buzd84}. Two recent
experiments reported an influence of the domain state on the transport properties of the
superconductor. In Nb/Co bilayers it was found that the superconducting transition
temperature T$_c$ and critical current density J$_c$ (defined using a voltage criterion)
are higher at the coercive field of the Co layer \cite{kinsey01}; a similar effect on
$J_c$ was observed in Co/Nb/Co trilayers, again using a voltage criterion \cite{koba03}.
Here we present results of a similar study, which differs in two ways from the ones cited
above. We use F/S/F trilayers with S = Nb and F = Cu$_{0.46}$Ni$_{0.54}$, which is a very
weak ferromagnet with a saturation magnetization $M_s$ of about 0.1~$\mu_B$ ($\mu_B$ is
the Bohr magneton) \cite{rusa02} to be compared to about 1~$\mu_B$ for Co. The experiment
we perform is measuring the depairing current by a pulsed-current method. This quantity
reflects the superconducting order parameter directly, and we showed before that it can
be used in S/F systems \cite{geers01}. We find that at least close to T$_c$ the critical
current of the trilayer as function of an in-plane magnetic field peaks at the coercive
field H$_{co}$ of the F-layer, which means that superconductivity is enhanced in the
domain state.

\section{Sample preparation and magnetic properties}
\begin{figure}
\begin{center}
\includegraphics[width=7cm]{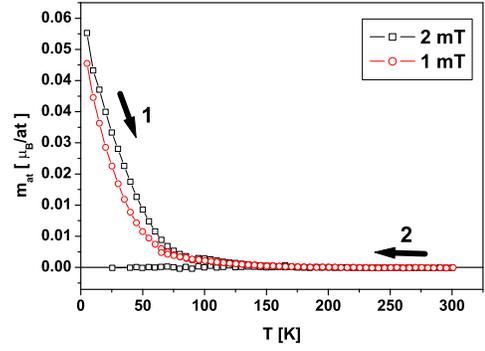}
\end{center}
\caption{Magnetization $M$ as a function of temperature $T$ for a CuNi/Nb/CuNi trilayer
after saturation at 5~K in the fields indicated. Arrows denote the measurement sequence.
The Curie temperature is around 110 K.}
\end{figure}
Single F-films and F/S/F trilayers were DC-magnetron sputtered in an ultra high vacuum
system with base pressure of about 10$^{-9}$ mbar and sputtering argon pressure about
6*10$^{-3}$ mbar. A Cu$_{1-x}$Ni$_x$ target was used with $x$ = 0.50, which yielded a
slightly different Ni concentration in the samples of $x$ =0.54. We will call this alloy
CuNi. The thickness of the F-layers was chosen to be 9~nm, which is about 0.5~$\xi_F$
\cite{ryaz01}. The S-layer thickness was chosen at 23~nm in order to have the
superconducting transition temperature around 4~K \cite{rusa02}.
\begin{figure}
\begin{center}
\includegraphics[width=7cm]{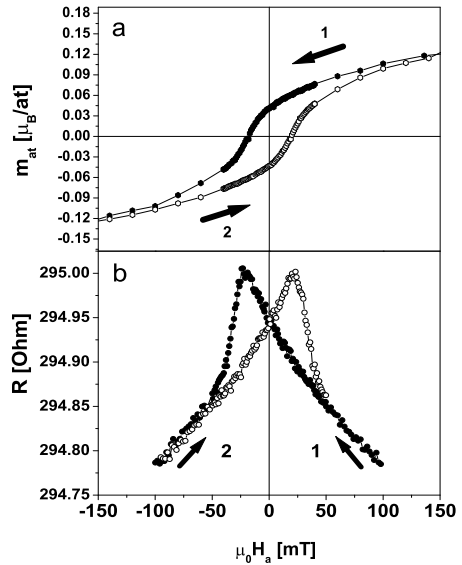}
\end{center}
\caption{Dependence of (a) magnetization $M$ and (b) resistance $R$ on magnetic field
H$_a$ for a CuNi/Nb/CuNi trilayer at 5 K. Filled symbols are used for the field sweep
from positive to negative field (denoted 1), open symbols for the opposite direction
(denoted 2). The coercive  field H$_c$ is around 22 mT.}
\end{figure}
Electron beam lithography was used to pattern the samples for 4-point measurements, with
the bridge between the voltage contacts 2~$\mu$m wide and 20~$\mu$m long. The patterned
samples were Ar-ion etched. Unstructured samples were measured by SQUID-magnetometry in
order to determine M$_s$, H$_{co}$ and the Curie temperature T$_{Curie}$. Fig.~1 shows
the typical dependence of the magnetization $M$ on temperature $T$ for a single F-layer,
using a standard procedure: the sample was magnetized at 5~K to its saturation in a field
of 0.7~T, then the field was set to a small value (typically of about 0.1~mT), and $M$(T)
was measured up to 300~K and back down to 5~K. T$_{Curie}$ was defined at the temperature
where irreversibility sets in when cooling down. The dependence of $M$ on applied field
$H_a$ at 5~K of a single F-layer is given in Fig.~2a. It shows a hysteresis loop typical
for a ferromagnet with H$_{co}$ about 25~mT. To determine possible differences in
H$_{co}$ between structured and unstructured samples caused by the sample shape
anisotropy, we measured the magnetoresistance R(H$_a$) of a structured samples at 5~K.
With H$_a$ in the plane of the film, but perpendicular to the bridge and therefore the
current, R(H$_a$) behaves as shown in Fig. 2b. Two peaks are found at H$_{co}$, due to
the anisotropic magnetoresistance effect.  The slight difference in aspect ratio $A$ for
the structured ($A$ = 10) and the unstructured ($A$ = 5) samples does not lead to a
significant difference in H$_{co}$.

\section{Superconducting properties and discussion}
\begin{figure}
\begin{center}
\includegraphics[width=7cm]{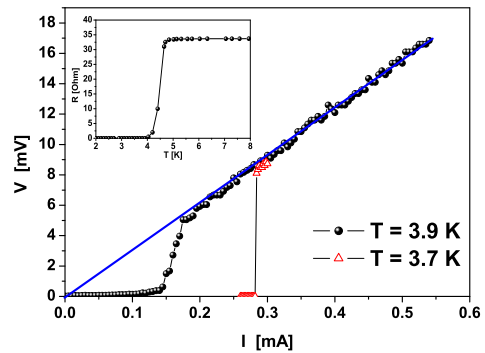}
\end{center}
\caption{Current (I) versus voltage (V) characteristics of a CuNi/Nb/CuNi trilayer shaped
as 2 $\mu$m wide bridges at the temperatures indicated. The line shows the normal (ohmic)
resistance.  The transition from the normal to the superconducting state, shown in the
inset, occurs at T$_c$ = 4.0 K.}
\end{figure}
For the determination of the critical current I$_c$ at different temperatures a
pulsed-current method was used. Current pulses of about 3~ms with growing amplitude were
sent through the sample below T$_c$~= 4.0~K (shown in  the onset of Fig.~3). Each pulse
was followed by a long pause of about 7~s. The voltage response of the system was
observed on a oscilloscope triggered for the time of a single pulse. To improve the
signal resolution a differential amplifier combined with low-noise band filters was used.
Two typical current (I)- voltage (V) characteristics are shown in Fig.~3, taken at
temperatures 3.9~K (reduced temperature t = T / T$_c$ = 0.98) and 3.7~K (t = 0.93). One
can see a clear jump from the superconducting into the normal state at I$_c$, which we
designate the depairing current \cite{geers01}. For all samples, down to t = 0.98 a small
onset voltage was observed below I$_c$, probably because of vortex motion. In order to be
sure that this has no influence on the depairing current, the sample temperature was
probed during every current pulse. A significant increase was found only when I$_c$ was
reached. \\
\begin{figure}
\begin{center}
\includegraphics[width=7cm]{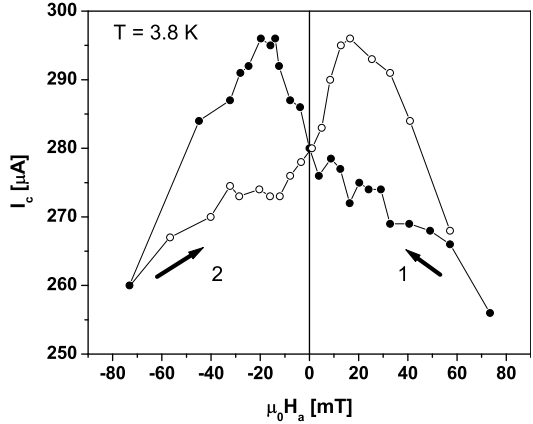}
\end{center}
\caption{Depairing current I$_c$ of the CuNi/Nb/CuNi trilayer as  function of applied
magnetic field H$_a$ at T = 3.8 K. Filled symbols are used for the field sweep from
positive to negative field (denoted 1), open symbols for the opposite direction (denoted
2).}
\end{figure}
The dependence of I$_c$ as function of H$_a$ at T = 3.8 K is shown in Fig.~4. Upon
lowering H$_a$ from the positive high field side, I$_c$  rises and reaches its maximum
value at a negative value of about -20~mT, quite close to H$_{co}$. It then decreases
again. When increasing H$_a$ from a large negative field, a similar maximum occurs at
+20~mT, with clearly hysteretic behavior. Clearly, the suppression of superconductivity
in the S-layer by the proximity effect is less for the case when a domain structure is
present in the outer F-layers, which is similar to the findings in
refs.~\cite{kinsey01,koba03}. Note that the result does not imply the existence of a
coupling between the F-layers : the variation of exchange field directions in the domain
walls might be sufficient to reduce the suppression. The scatter in the data points in
Fig.~4 may be caused by the changing domain structure; either a varying suppression of
the order parameter or an induced vortex state \cite{ryaz02} could result in a slightly
different current distribution at each point. The main result of a peak in I$_c$ around
H$_{co}$ is also found for the case of thick ($\approx$~50~nm) ferromagnetic layers. This
seems to exclude the possibility that effects of inducing an inhomogeneous order
parameter in the F-banks play a role, at least in this temperature regime.
\section{Acknowledgements}
This work is part of the research program of the 'Stichting voor Fundamenteel Onderzoek
der Materie (FOM)', which is financially supported by NWO.

\end{document}